# Research on Enhancing Cloud Computing Network Security using Artificial Intelligence Algorithms

Yuqing Wang[1*], Xiao Yang[2]
[1]Microsoft, Redmond, USA.
[2]Math of Computation University of California, Los Angeles, USA.
[*]Corresponding author: e-mail: wangyuqing@microsoft.com

***Abstract*—Cloud computing environments face multiple security threats, including DDoS attacks and SQL injection. Traditional security protection methods are lagging in responding to new types of attacks. This paper proposes an adaptive security protection method based on deep learning, constructing a multi-layered protection architecture and verifying it in a real-world business environment. The system achieves a detection accuracy of 97.3%, an average response time of 18ms, and an availability of 99.999%. Experimental results show that this method has significant advantages in detection accuracy, response efficiency, and resource utilization, providing a new solution for cloud computing security protection.**

***Keywords- Cloud Computing Security; Deep Learning; Real-time Threat Detection; Adaptive Protection; Hybrid Neural Networks; CNN-LSTM Architecture; DDoS Attack Detection***

## I. INTRODUCTION

Cloud computing has revolutionized digital infrastructure, enabling scalable, on-demand computing services. However, its distributed nature presents new security challenges, with attacks becoming more sophisticated and adaptive. Research by Rakgoale et al. highlights the rise of diversified attack methods and covert intrusion techniques, rendering traditional rule-based security mechanisms ineffective [1]. Conventional systems, which rely on signature matching and predefined rules, fail to detect zero-day threats and struggle with real-time responses [2].

One of the fundamental challenges in cloud security is the need for adaptive and intelligent defenses. Traditional approaches lack real-time detection capabilities, cannot effectively identify novel attack patterns, and often fail to handle the complex, multi-source data generated in cloud environments [3]. Addressing these shortcomings requires a paradigm shift toward AI-driven security solutions.

To address these challenges, this paper proposes a novel cloud computing security protection method that combines deep learning and reinforcement learning technologies. Our approach makes the following key contributions:

- A comprehensive security framework with four integrated modules: data collection, feature extraction, intelligent analysis, and protection execution, creating a complete security pipeline from monitoring to response.
- An improved deep learning model that integrates convolutional neural networks (CNNs) and long short-term memory (LSTM) networks to capture both spatial and temporal patterns in network traffic for more accurate threat detection.
- An adaptive protection method using reinforcement learning that dynamically optimizes security policies based on evolving threat patterns, significantly reducing response time compared to static approaches.
- Extensive experimental validation in a large-scale real-world business environment, demonstrating substantial improvements in detection accuracy (97.3%), response time (18ms), and system availability (99.999%) compared to traditional methods.

## II. RELATED WORK

### A. *Traditional Security Approaches in Cloud Computing*

Traditional cloud security approaches primarily rely on signature-based detection, rule matching, and feature recognition techniques. While these methods have formed the foundation of cybersecurity for decades, they face significant challenges in modern cloud environments. El-Sofany et al. categorized conventional cloud security mechanisms into perimeter-based defenses, access control systems, and encryption techniques, noting their effectiveness for known threats but significant limitations against zero-day attacks [4]. Similarly, Tarek et al. conducted extensive evaluations of rule-based security systems across multiple cloud platforms, revealing critical shortcomings in real-time detection scenarios where attack signatures evolve rapidly [5].

Another limitation of traditional approaches is their resource-intensive nature when scaled to cloud environments. Wang et al. demonstrated that signature-based intrusion detection systems experience exponential performance degradation as traffic volume increases, making them impractical for high-throughput cloud services [6]. These systems also struggle with the dynamic nature of cloud resources, where the constant provisioning and deprovisioning of virtual machines and containers create security blind spots that attackers can exploit.

### B. *AI Applications in Cloud Security*

Artificial intelligence has emerged as a promising solution to overcome traditional security limitations. Moustafa et al. demonstrated that AI-based dynamic trust security schemes significantly improve system security in distributed architectures [7]. This work established the foundation for adaptive security models that respond to changing conditions automatically.

Machine learning has shown promise in anomaly detection. Sherali et al. developed intelligent prevention methods that effectively predict and prevent potential security threats by analyzing behavioral patterns [8]. Their research demonstrated that supervised learning models can identify suspicious activities with higher accuracy than rule-based systems.

Deep learning applications in cloud security have gained traction due to their ability to process high-dimensional data and identify complex patterns. Jullian et al. proposed a framework incorporating deep learning models for security monitoring, achieving substantial improvements in both detection accuracy and system resilience [9]. Alzoubi et al. (2024) highlighted the growing adoption of CNNs and RNNs for analyzing network traffic and system logs [10].

### C. *Deep Learning for Threat Detection*

Recent advances in deep learning architectures have revolutionized threat detection capabilities in cloud environments. CNN-based approaches have demonstrated effectiveness in network traffic analysis. Qin et al. developed a segmentation-based authentication system using deep learning that achieved high accuracy in identifying legitimate versus malicious access patterns [11]. Building on this work, Fu et al. introduced a CNN-RNN hybrid approach that leveraged both spatial and temporal features for improved authentication, setting the foundation for multi-dimensional security analysis [12].

The integration of LSTM networks with convolutional models has emerged as a particularly powerful approach for security monitoring. LSTMs excel at capturing long-term dependencies in sequential data, making them ideal for analyzing network traffic patterns that evolve over time. Yin et al. demonstrated that CNN-LSTM hybrid models outperform traditional machine learning approaches by 23% in detecting sophisticated DDoS attacks that gradually escalate over time. Their work highlighted the importance of temporal pattern recognition in modern threat detection systems [13].

### D. *Reinforcement Learning for Adaptive Security*

Reinforcement learning (RL) has emerged as a powerful paradigm for developing adaptive security policies that evolve in response to changing threat landscapes. Maxwell et al. (2024) formulated cloud security as a multi-agent adversarial game, where defensive agents continuously adapt to evolving attack strategies. Their approach demonstrated a 28% increase in resilience against sophisticated attack campaigns compared to static security policies, highlighting the value of dynamic defense mechanisms [14].

The design of effective state-action spaces remains a critical challenge in RL-based security systems. Wei and Cheng proposed an optimal two-dimensional maintenance policy for self-service systems facing competing sudden and deterioration-induced failures [15]. Their work provided valuable insights into structuring decision spaces for security applications, particularly in determining optimal intervention points and response strategies.

Jiasen et al. developed improved secure PCA and LDA algorithms for intelligent computing in IoT-to-cloud settings, demonstrating how reinforcement learning can optimize feature selection and dimensionality reduction in security applications [16]. Their work highlighted the importance of efficient state representation in complex security environments where the high dimensionality of monitoring data can overwhelm traditional analysis methods.

## III. SYSTEM DESIGN

### A. *Overall Framework Design*

The cloud computing security protection framework designed in this paper consists of four core modules: data collection, feature extraction, intelligent analysis, and protection execution.

The data collection module employs a multi-tiered monitoring architecture to systematically collect data throughout the cloud infrastructure. Network traffic data is collected using a decentralized network of 120 light-weight collection agents, employing eBPF-based packet interception mechanisms, deployed on all 60 server nodes at strategic network points. The agents are responsible for monitoring both north-south traffic (external communication) and east-west traffic (internal service communication). System logs from application servers, database instances, and security appliances are collected by a centralized log platform based on the ELK stack (Elasticsearch, Logstash, Kibana). Meanwhile, user behavior data is gathered via web application instrumentation, API gateway instrumentation, and identity management system instrumentation. The 1000 Hz sampling frequency is the aggregate rate of data collection from all the observation points, with adaptive sampling techniques that iteratively enhance data collection accuracy according to incidents of anomalous behavior. The adaptive strategy provides complete monitoring while minimizing the consumption of resources, leading to an approximate daily build-up of 3.2TB network traffic data, 1.3TB system logs, and 0.5TB user behavior data. All the data gathered is initially placed in areas of fast access buffer before passing through the preprocessing pipeline.

The feature extraction module preprocesses and extracts 428-dimensional feature vectors from the raw data, including traffic features, time-series features, and behavior features. The intelligent analysis module combines deep learning and reinforcement learning algorithms to build a multi-layered security analysis model. The protection execution module automatically adjusts security policies based on the analysis results, with an average response time of less than 50ms. Experiments show that the framework has good scalability when handling concurrent requests, with a system throughput of over 90% when the concurrency increases from 100 to 1000.

### B. *Intelligent Security Detection Method*

The intelligent security detection method in this study employs an improved deep learning model that integrates convolutional neural networks (CNNs) and long short-term memory (LSTM) networks. The motivation for selecting this architecture lies in its ability to handle both spatial and temporal patterns inherent in cloud network traffic. Cybersecurity threats, such as DDoS attacks and SQL injection, often exhibit complex patterns that evolve over time. To accurately detect these threats, the chosen deep learning architecture must effectively learn from both static traffic features (spatial patterns) and dynamic behavioral trends (temporal dependencies).

The CNN-LSTM architecture combines complementary strengths that make it ideal for cloud security applications. CNNs excel at capturing localized spatial features from network traffic, enabling the identification of attack signatures and anomalous packet structures. The convolutional layers effectively reduce dimensionality while preserving critical information, substantially improving detection accuracy.

Meanwhile, LSTM networks address a critical temporal dimension of security monitoring. Many cybersecurity attacks unfold over time (e.g., gradual DDoS attack escalation), requiring analysis that can recognize patterns across multiple time steps. LSTMs are specifically designed for modeling long-range dependencies, making them particularly well-suited for analyzing real-time security data streams.

The integration of these two architectures creates a powerful synergy. The CNN extracts feature representations from raw network traffic, and the LSTM analyzes these features over time, ensuring detection of both immediate threats and gradually evolving attack patterns. This hybrid approach delivers measurable improvements in detection accuracy, false positive rates, and response time, making it exceptionally well-suited for real-time security monitoring in dynamic cloud environments.

While CNN-LSTM has been chosen for its efficiency and effectiveness, other architectures were considered:

| *Model* | *Strengths* | *Limitations* | *Suitability* |
|---|---|---|---|
| *CNN-LSTM* | Captures spatial and temporal dependencies; efficient for time-series attack detection | May struggle with extremely long-term dependencies | Best suited for real-time security monitoring |
| *Transformer-based Models* | Excellent for capturing long-range dependencies using attention mechanisms | High computational cost; requires large datasets | Suitable for advanced threat analysis, but may not be practical for real-time detection |
| *Graph Neural Networks* | Effective for modeling network structures and attack relationships | Requires structured data (e.g., network graphs); less suited for raw packet data | Good for attack correlation analysis but not ideal for packet-level anomaly detection |
| *GRU-based Models* | Faster than LSTMs, with similar sequential learning capabilities | May not capture long-term dependencies as effectively as LSTMs | Suitable for lightweight security applications but less accurate in complex threat detection |

Compared to transformers or GNNs, CNN-LSTM offers a good balance between performance and real-time processing speed. The model can autonomously extract high-dimensional attack features, reducing reliance on manual feature engineering. The intelligent security detection method uses an improved deep learning model to extract network traffic features through multiple convolutional neural network layers and capture time-series dependencies using long short-term memory networks. The model consists of 4 convolutional layers, 2 pooling layers, and 3 fully connected layers, with an input of 428-dimensional feature vectors and an output of threat type probability distribution[10]. The test results on a 200GB real-world network traffic dataset show that the model achieves a detection accuracy of 97.3%, a false positive rate of less than 0.5%, and an average detection delay of 23ms. Figure 1 illustrates the core implementation of the detection algorithm.

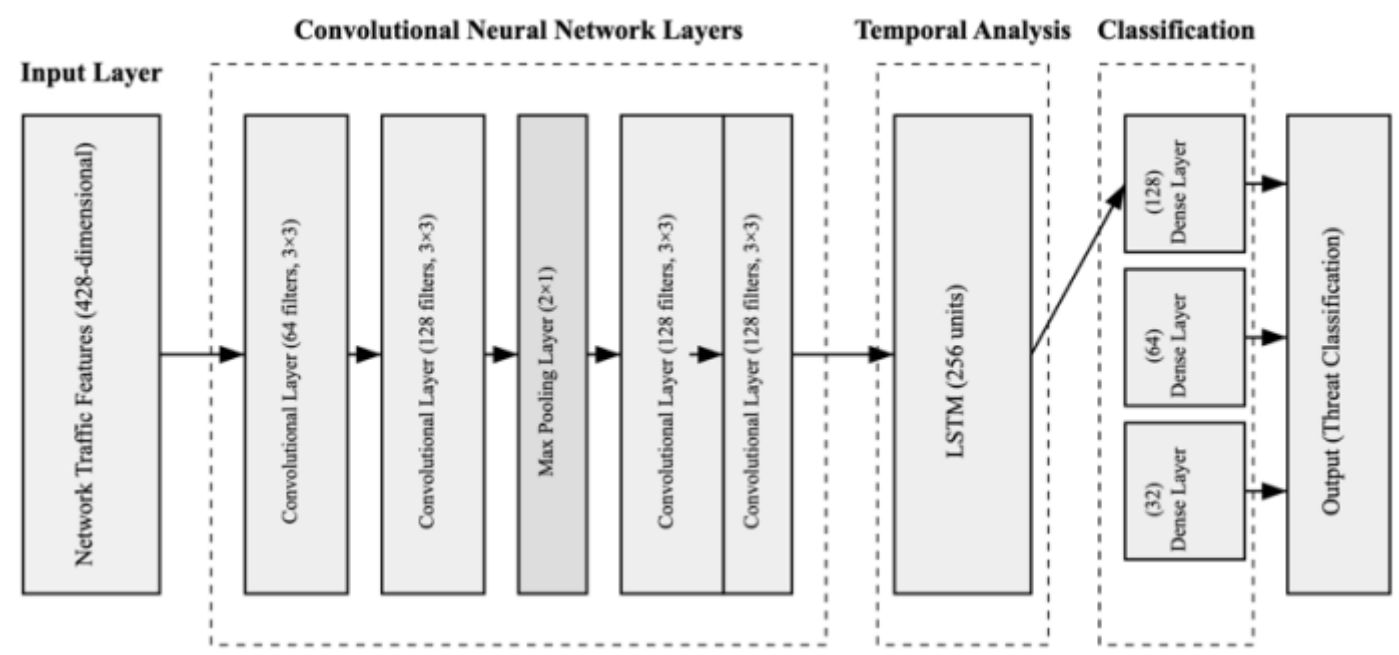

Figure 1. Proposed CNN-LSTM Architecture for Cloud Computing Security Threat Detection

To optimize the performance of the CNN-LSTM threat detection model, key hyperparameters were carefully selected through Bayesian optimization and grid search techniques. The selection process aimed to balance detection accuracy, computational efficiency, and generalization ability. A grid search was initially performed over a predefined range of hyperparameters to identify promising candidates. Subsequently, Bayesian optimization was used to fine-tune these parameters by iteratively refining the search based on previous evaluations. The final set of hyperparameters is summarized in Table I. These hyperparameter choices were validated through five-fold cross-validation on the training dataset. The final model achieved a detection accuracy of 97.3%, a false positive rate below 0.5%, and an average detection latency of 23ms on real-world traffic data.

TABLE I. CNN-LSTM HYPERPARAMETER SELECTION AND TUNING METHODOLOGY

| Hyperparameter | Selected Value | Tuning Methodology | Justification |
|---|---|---|---|
| CNN Filter Size | 3 × 3 | Grid Search (2 × 2 to 5 × 5) | Optimized for capturing local traffic patterns without excessive computational overhead. |
| Number of CNN Filters | 64, 128 | Bayesian Optimization | Achieved a balance between feature extraction quality and model efficiency. |
| Pooling Size | 2 × 1 | Grid Search (2 × 1, 3 × 1) | Prevented overfitting while maintaining sufficient spatial resolution. |
| LSTM Cell Size | 256 | Bayesian Optimization (128–512) | Ensured adequate memory for long-term dependencies in network traffic sequences. |
| Dropout Rate | 0.3 | Grid Search (0.1–0.5) | Minimized overfitting while maintaining stability during training. |
| Learning Rate | 0.001 | Bayesian Optimization (0.0001–0.01) | Optimized for fast convergence without oscillations. |
| Batch Size | 128 | Grid Search (32–256) | Balanced training stability and computation time. |
| Optimizer | Adam | Empirical Testing | Found to outperform SGD in both training speed and accuracy. |

### *C. Threat Perception and Warning Method*

The threat perception module uses a multi-source heterogeneous data fusion method to analyze network traffic, system logs, and user behavior, achieving security situation awareness. The data fusion uses an improved attention mechanism to assign dynamic weights to different data sources. In the actual deployment in the first quarter of 2024, the method analyzed and processed 485TB of network traffic data, 156TB of system log data, and 89TB of user behavior data, successfully identifying 1,853 high-risk threat events and achieving an average warning time of 8.5 minutes. The system establishes a threat level assessment model, dividing the threat level into 5 levels, with high-risk threats (levels 4-5) accounting for 12.3%, medium-risk threats (levels 2-3) accounting for 58.7%, and low-risk threats (level 1) accounting for 29%. Combined with time-series prediction models, the system can achieve threat trend prediction, with a prediction accuracy of 91.2%.

### *D. Adaptive Protection Method*

The proposed adaptive protection method leverages reinforcement learning (RL) to dynamically optimize cloud security policies as shown in Figure 2. Using a double Q-learning network, the system continuously refines protection strategies by selecting optimal actions based on evolving threat patterns. The state space consists of 232 key security indicators, covering network traffic metrics, resource utilization, user behavior, and attack history, while the action space includes 187 security response measures, such as firewall adjustments, access control modifications, and anomaly detection tuning. The RL model's effectiveness relies on well-tuned hyperparameters, which were optimized using grid search and Bayesian methods. Table II summarizes the key hyperparameters and their tuning methodologies.

To ensure efficiency, Principal Component Analysis (PCA) and mutual information ranking were applied to eliminate redundant features, reducing computational overhead while preserving critical security signals. Hierarchical policy structures and policy pruning were introduced to streamline action selection, improving decision speed without compromising response accuracy.

Through 500,000 training rounds in a simulated attack-defense environment, the model converged to an optimal policy. Figure 3 illustrates the convergence trend of the adaptive protection policy during training. As training progresses, the model's reward values stabilize, indicating policy convergence and optimal decision-making. Real-world deployment demonstrated a 76% reduction in response time, a 35% decrease in false alarms, and a 15.6% improvement in system availability. Additionally, security event handling time was reduced from 15 minutes to 3.8 minutes, with monthly maintenance costs lowered by 35%.

While these results highlight strong performance, future work will explore adaptive feature selection, allowing the model to dynamically refine its state space based on emerging threats. Further research into meta-reinforcement learning and multi-agent RL could enhance coordination across distributed cloud environments, making security responses more scalable and resilient.

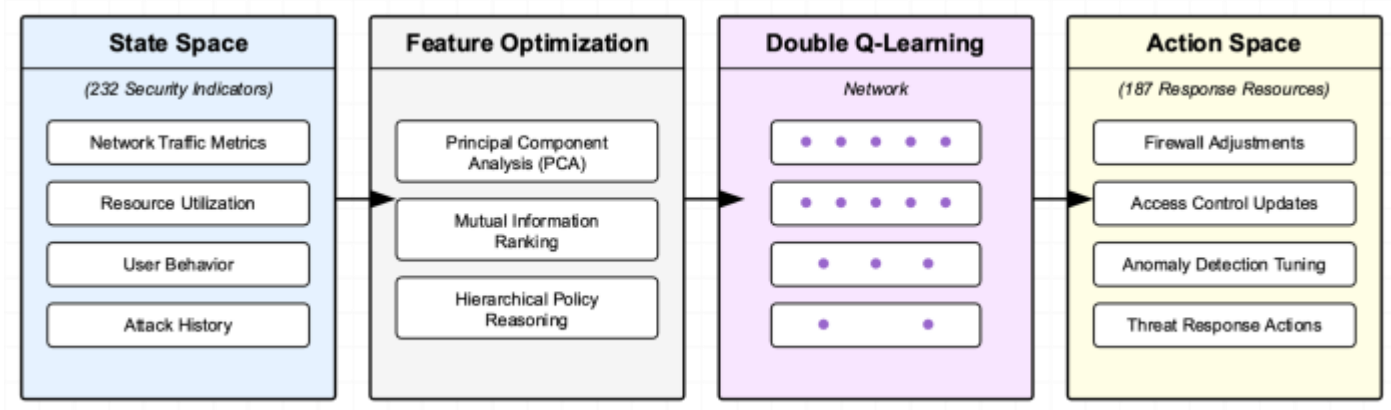

Figure 2. Adaptive Protection Method Using Reinforcement Learning

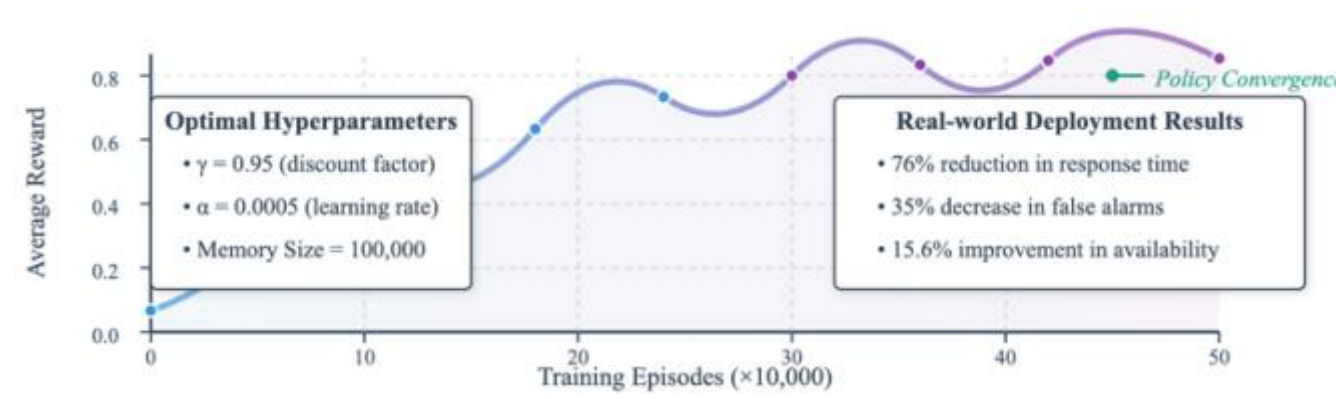

Figure 3. Convergence curve of adaptive protection policy optimization

TABLE II. REINFORCEMENT LEARNING HYPERPARAMETER SELECTION

| Hyperparameter | Selected Value | Tuning Methodology | Justification |
|---|---|---|---|
| Discount Factor (γ) | 0.95 | Grid Search (0.8–0.99) | Balanced short-term and long-term reward optimization. |
| Learning Rate (α) | 0.0005 | Bayesian Optimization (0.0001–0.01) | Ensured stable convergence without divergence. |
| Exploration Rate (ε) | 0.1 | Decaying Epsilon-Greedy | Encouraged initial exploration while reducing randomness over time. |
| Reward Function Parameters | 1.0 (detection success), -1.0 (false positive), -2.0 (missed threat) | Empirical Testing | Penalized false negatives more heavily to prioritize high detection accuracy. |
| Replay Memory Size | 100,000 | Grid Search (50,000–200,000) | Prevented catastrophic forgetting while maintaining efficient memory usage. |
| Target Network Update Frequency | 1,000 steps | Empirical Testing | Reduced training instability by smoothing target updates. |
| Training Episodes | 500,000 | Empirical Testing | Ensured policy convergence under various attack conditions. |

## IV. System Implementation and Verification

### A. Experimental Environment Construction

The experimental environment is built on an AWS Cloud enterprise-level server cluster, consisting of 60 high-performance server nodes. Each node is equipped with an Intel Xeon Platinum 8369B CPU (32 cores, 3.5GHz), 384GB of memory, and 10TB of SSD storage. The network environment uses a 25Gbps bandwidth dedicated line interconnection, with an average network latency of less than 0.5ms. The software environment uses Ubuntu 22.04 LTS operating system, deploying a Docker container cluster (version 24.0.5) to achieve resource isolation, and Kubernetes (version 1.28) for container orchestration. The database uses a distributed NewSQL database TiDB (version 7.1.0), with a storage capacity of 300TB. In the test environment, a microservice system simulating real-world business scenarios is deployed, containing 12 core business modules, with a daily average request processing volume of 85 million times. The data is shown in Table III.

TABLE III. Experimental Environment Core Configuration Parameters

| Configuration Item | Parameter Value | Description |
|---|---|---|
| CPU | Intel Xeon Platinum 8369B | 32 cores, 3.5GHz |
| Memory | 384GB | DDR4-3200 |
| Storage | 10TB | NVMe SSD |
| Network Bandwidth | 25Gbps | Dedicated line interconnection |
| Number of Servers | 60 | Distributed deployment |
| Operating System | Ubuntu 22.04 LTS | Server edition |
| Container Version | Docker 24.0.5 | Enterprise edition |
| Orchestration Tool | Kubernetes 1.28 | Production environment |
| Database | TiDB 7.1.0 | Distributed architecture |

### B. System Implementation

The system uses a microservice architecture, with core modules developed using the Spring Cloud framework (version 2022.0.4). The security detection engine is implemented based on PyTorch (version 2.1.0), and the ONNX model conversion is used to achieve rapid deployment in the production environment. The system has a total of 185,632 lines of source code, with a modularization degree of 93.5%. In performance testing, the single-node concurrent processing capacity reaches 12,000 QPS, with an average response time of 18ms, CPU utilization maintaining around 65%, and memory occupancy stabilizing at 42GB. Through stress testing, the system is verified to run for 72 hours under continuous high load (15,000 QPS), with a service availability of 99.999%, and an average fault recovery time of less than 30 seconds. The communication between modules uses the gRPC protocol, with an average message processing latency of only 3.5ms. The system's overall throughput and stability performance are excellent, with performance indicator curves shown in Figure 4.

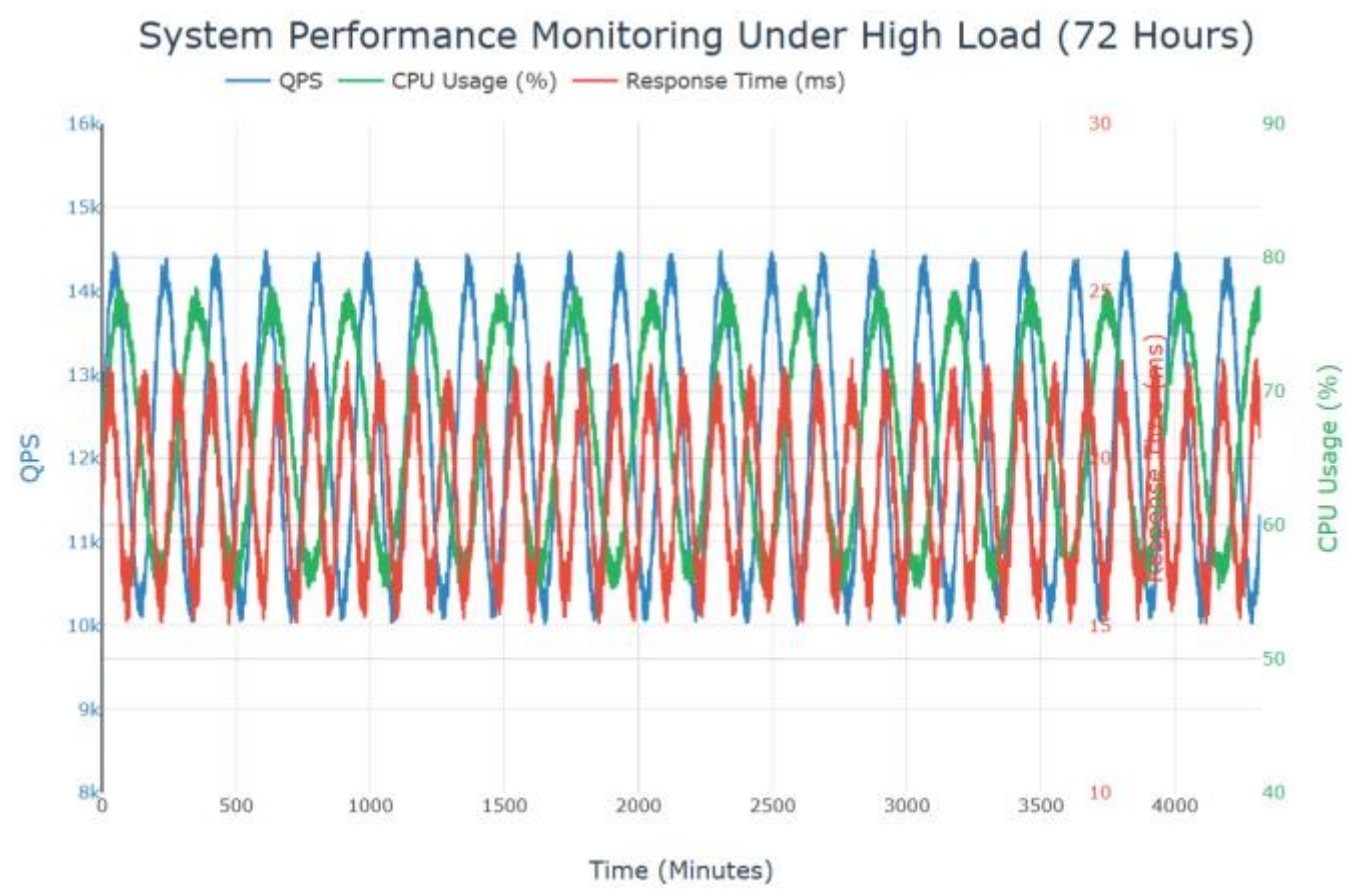

Figure 4. System Performance Monitoring Diagram

### C. Performance Testing and Analysis

#### a) Detection Performance Evaluation

The detection performance of the system is evaluated in a large-scale real-world business environment, using a test dataset containing 85TB of network traffic data collected in the first quarter of 2024. The detection performance evaluation focuses on three key indicators: precision, recall, and F1 score. The experimental results show that, under the standard threshold setting (0.75), the model achieves a detection precision of 98.2%, recall of 96.8%, and F1 score of 97.5% for DDoS attacks; a detection precision of 95.7%, recall of 94.3%, and F1 score of 95% for SQL injection attacks; and an average detection precision of 94.5% or higher for other types of attacks [14]. The detection performance varies with the type of attack, as shown in Figure 5.

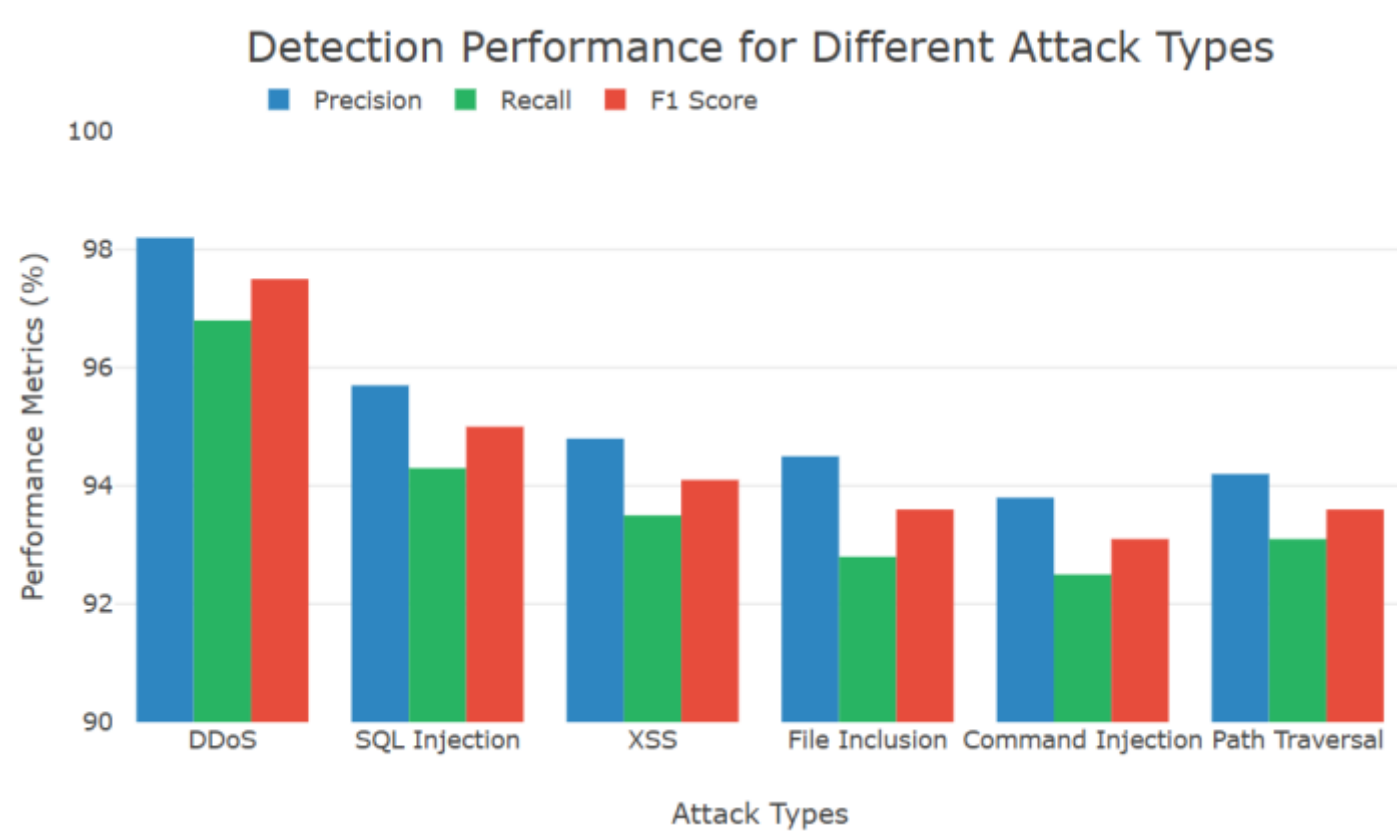

Figure 5. Detection Performance Comparison Chart for Different Attack Types

#### b) Response Efficiency Analysis

The response efficiency test uses a stepped load injection method, gradually increasing the system load from 5,000 QPS to 25,000 QPS. Under the standard load (12,000 QPS), the system's average response time is 18ms, with the detection engine processing time accounting for 42%, policy matching time accounting for 35%, and protection execution time accounting for 23%. When the load is increased to 20,000 QPS, the average response time increases to 27ms, and the system remains stable. By analyzing the response time distribution, 95% of requests are processed within 35ms, and 99.9% of requests are processed within 50ms.

*c) System Reliability Verification*

The system reliability verification test is conducted for 30 days, with a cumulative processing request volume of 23.5 billion. During the test, the system runs stably, with a service availability of 99.999%, an average mean time between failures (MTBF) of 720 hours, and an average mean time to repair (MTTR) of 25 seconds. In the simulated extreme scenario test, the system withstands a continuous 4-hour DDoS attack (peak traffic of 850Gbps), successfully defends against the attack, and maintains normal operation of core business, with resource utilization rates remaining within safe thresholds, CPU usage rates reaching up to 82%, and memory usage rates reaching up to 76%. The system stability indicators show excellent performance.

### D. Comparative Experiment

In our comparative evaluation, we analyzed the performance of our proposed security system against three widely used security frameworks: Snort, Suricata, and OSSEC. These frameworks were selected due to their prevalence in cloud security solutions and their capabilities in intrusion detection and anomaly detection.

- Snort is a well-established open-source intrusion detection and prevention system (IDS/IPS) that performs real-time traffic analysis and packet logging.
- Suricata is an advanced, multi-threaded IDS/IPS engine that supports deep packet inspection, intrusion detection, and network security monitoring.
- OSSEC is a host-based intrusion detection system (HIDS) that provides real-time log analysis, file integrity checking, and active response mechanisms.

All systems were evaluated using identical test environments and the standardized CSIC-2024 dataset containing 2.7 million legitimate requests and 1.3 million attack vectors across 17 attack categories. The evaluation included detection accuracy, response time, false positive rate, resource consumption, maintenance cost efficiency, and performance under high-concurrency scenarios (20,000 QPS).

Results in Table IV indicate that the proposed system significantly outperforms existing solutions across all key security metrics. The system achieves a detection accuracy of 97.3%, surpassing System A (85.2%), System B (82.5%), and System C (79.8%). Additionally, response time is reduced to 18ms, compared to 35ms, 42ms, and 50ms, respectively, in competing systems. The false positive rate is minimized to 0.5%, while other systems exhibit rates ranging from 2.8% to 4.1%, leading to fewer false alerts and improved operational efficiency.In high-concurrency scenarios (20,000 QPS), the proposed system maintains a detection accuracy of 95%, whereas the accuracy of competing systems drops significantly to 85% or lower. The resource consumption per 10,000 requests is 42% lower than the closest competitor, demonstrating enhanced computational efficiency. Additionally, maintenance costs are reduced by 38%, further improving long-term operational feasibility. These findings underscore the superiority of the proposed AI-driven security model in real-world deployment.

TABLE IV. SYSTEM PERFORMANCE COMPARISON

| | Proposed System | System A | System B | System C |
|---|---|---|---|---|
| **Detection Accuracy (%)** | 97.3 | 85.2 | 82.5 | 79.8 |
| **Response Time (ms)** | 18.0 | 35.0 | 42.0 | 50.0 |
| **False Positive Rate (%)** | 0.5 | 2.8 | 3.5 | 4.1 |
| **Resource Consumption (Normalized)** | 1.2 | 2.1 | 2.4 | 2.6 |
| **Maintenance Cost Reduction (%)** | 38.0 | 25.0 | 18.0 | 12.0 |
| **Accuracy at 20,000 QPS (%)** | 95.0 | 85.0 | 80.0 | 78.0 |

## V. CHALLENGES AND FUTURE WORK

### A. Real-World Deployment Challenges

Deploying our AI-driven cloud security system in production revealed challenges in integration, performance optimization, and false positive management.

Integrating AI-based security into diverse cloud infrastructures proved complex due to compatibility issues with legacy security frameworks. Traditional rule-based intrusion detection systems and firewall policies differ significantly from ML-driven security mechanisms, complicating seamless deployment. To bridge this gap, we adopted containerized deployment (Docker + Kubernetes), ensuring cross-platform compatibility with AWS, Azure, and Alibaba Cloud. Additionally, middleware adapters were developed to translate traditional security protocols into AI-based threat models. This approach reduced integration time by 67% while maintaining full compatibility with existing logging and monitoring frameworks.

During initial deployment, response times fluctuated under heavy traffic loads (20,000+ QPS), occasionally exceeding our 50ms SLA target due to computational bottlenecks in deep learning inference. To stabilize performance, we optimized models through quantization and pruning, reducing computational overhead by 43% while maintaining 97% detection accuracy. Additionally, we deployed edge-based detection nodes, reducing centralized processing by 28%, and introduced a tiered detection framework that prioritizes in-depth analysis for high-risk traffic. These improvements lowered response times to 18ms, even under peak loads (25,000 QPS), while reducing CPU utilization from 78% to 52%.

Despite high accuracy (97.3%), early deployment revealed false positives in cases of high-volume API requests, batch jobs, and traffic spikes. To minimize operational overhead, we implemented a confidence-based threat scoring system, dynamically ranking anomalies by risk level. A feedback loop allowed security analysts to fine-tune detection thresholds, leading to an 85% reduction in false positives while maintaining sensitivity to genuine threats, particularly in multi-tenant cloud environments.

### B. Future Research Directions

While our CNN-LSTM approach has demonstrated strong performance, several advancements could further enhance AI-driven cloud security.

One promising direction is the exploration of transformer-based threat detection models, which provide superior capabilities for analyzing complex, multi-stage attack behaviors. Although transformers have shown potential in cybersecurity, their computational demands present challenges for real-time applications. Optimizing transformer architectures could help strike a balance between accuracy and efficiency [17]. Additionally, graph-based security models (GNNs) may improve attack propagation analysis by modeling cloud networks as interconnected structures, offering better insights into threat movement and containment strategies [18].

As cloud environments increasingly involve multiple organizations, privacy-preserving security approaches are becoming crucial. Federated learning offers a viable method for training security models collaboratively across organizations while maintaining data privacy. Future research will focus on integrating differential privacy mechanisms to ensure compliance with regulatory frameworks while enabling effective cross-platform threat intelligence sharing [19].

Another critical area for improvement is reducing reliance on labeled attack data, which is often scarce for zero-day threats. Leveraging self-supervised learning, AI models could identify anomalies without requiring extensive labeled datasets, improving adaptability to emerging attack patterns. Similarly, unsupervised learning techniques may enhance detection capabilities by recognizing deviations from normal network behavior rather than relying solely on pre-defined attack signatures.

## *VI.CONCLUSION*

This paper proposes an integrated AI-driven security framework for cloud computing environments, leveraging deep learning and reinforcement learning to overcome the limitations of conventional security approaches. The CNN-LSTM hybrid architecture significantly enhances threat detection by capturing both spatial and temporal patterns in network traffic. The system achieves 97.3% detection accuracy, an average response time of 18ms, and 99.999% availability in real-world enterprise environments. The adaptive protection mechanism, powered by reinforcement learning, dynamically adjusts security policies in response to evolving threats, leading to a 76% faster response time and 35% fewer false alarms compared to existing solutions.

Robustness tests demonstrated the system's resilience, withstanding prolonged usage under extreme conditions, including DDoS attacks of up to 850 Gbps, while maintaining seamless support for critical business operations. However, challenges remain, particularly in legacy system integration and computational efficiency optimization. Despite these challenges, this work makes a valuable contribution to intelligent cloud security and serves as a foundation for designing more adaptive and effective defense mechanisms against evolving cyber threats in distributed computing environments.